\RequirePackage{ifpdf}
\documentclass{PoS}
\pdfoutput = 1
\usepackage{cite}
\usepackage{amsmath}  
\usepackage{amssymb} 
\usepackage{dsfont}
 \usepackage{multirow}
\usepackage{psfrag,scalefnt}
\usepackage{paralist}

\usepackage{color}
\let\normalcolor\relax

\newcommand{\vus}{|V_{us}|}
\newcommand{\vcb}{|V_{cb}|}

\newcommand{\vub}{|V_{ub}|}

\newcommand{\mev}{\, {\rm MeV}}

\newcommand{\be}{\begin{equation}}
\newcommand{\ee}{\end{equation}}
\newcommand{\bea}{\begin{eqnarray}}
\newcommand{\eea}{\end{eqnarray}}

\newcommand{\ba}{\begin{array}}
\newcommand{\ea}{\end{array}}

\def\kpn{K^+\rightarrow\pi^+\nu\bar\nu}

\def\klpn{K_{L}\rightarrow\pi^0\nu\bar\nu}

\def\ksm{K_S\to\mu^+\mu^-}
\newcommand{\BR}{{\cal B}}

\title{ Standard Model Predictions for Rare $K$ and $B$
  Decays \\ without $|V_{cb}|$ and $|V_{ub}|$
  Uncertainties\footnote{Submitted also to the DISCRETE2020-2021 Conference}}

\ShortTitle{Standard Model Predictions for Rare $K$ and $B$
  Decays}

\author{Andrzej J. Buras and \speaker{Elena Venturini}%
  \\
TUM-IAS, Lichtenbergstr. 2a, D-85748 Garching, Germany \\
Technical University Munich, Physics Department, D-85748 Garching, Germany,\\
E-mail: \email{aburas@ph.tum.de}, \email{elena.venturini@tum.de}}

  \abstract{The persistent tensions between inclusive and exclusive determinations
  of $|V_{cb}|$ and  $|V_{ub}|$ weaken the power of  theoretically
  clean rare $K$ and $B$ decays   in the search for new physics (NP).
  We demonstrate how this uncertainty can be practically removed by
  considering within the SM suitable ratios of various  branching ratios.
  This includes the branching ratios for $K^+\rightarrow\pi^+\nu\bar\nu$,
  $K_{L}\rightarrow\pi^0\nu\bar\nu$,
 $K_S\to\mu^+\mu^-$, $B_{s,d}\to\mu^+\mu^-$ and $B\to K(K^*)\nu\bar\nu$.
 Also $\varepsilon_K$, $\Delta M_d$,  $\Delta M_s$ and the mixing induced
 CP-asymmetry $S_{\psi K_S}$, all measured already  very precisely, play an important role
 in this analysis. The highlights of our analysis are 16 $|V_{cb}|$ and $|V_{ub}|$
 independent ratios that often are independent of the CKM parameters or
 depend only on the angles $\beta$ and $\gamma$ in the Unitarity Triangle
 with $\beta$ already precisely known and $\gamma$ to be measured precisely
 in the  coming years by the LHCb and Belle II colaborations. Once $\gamma$
 is measured precisely these 16 ratios taken together are expected to
 be a powerful tool in the search for new physics.
 Assuming no NP in  $|\varepsilon_K|$ and $S_{\psi K_S}$  we determine independently of  $|V_{cb}|$: $\mathcal{B}(K^+\rightarrow\pi^+\nu\bar\nu)_\text{SM}= (8.60\pm0.42)\times 10^{-11}$ and 
 $\mathcal{B}(K_L\rightarrow\pi^0\nu\bar\nu)_\text{SM}=(2.94\pm 0.15)\times 10^{-11}$. This are the most precise determinations to date. Assuming no NP in $\Delta M_{s,d}$ allows to
   obtain analogous results for all $B$ decay  branching ratios considered in our paper without any CKM uncertainties.
  
 }

\FullConference{CKM Workshop 2021, 22-26 November 2021, Melbourne, Australia
}

\begin{document}

\section{Introduction}
The rare $K$ and $B$ decays and the quark mixing being GIM suppressed in the Standard Model (SM) and simultaneously
being often theoretically clean are very powerful tools for the search of
New Physics (NP) \cite{Buras:2020xsm}.
Unfortunately the persistent tensions between inclusive and exclusive determinations of  $|V_{cb}|$ and $\vub$ (see e.g.
\cite{Bordone:2019guc,Bordone:2021oof,Ricciardi:2021shl,Aoki:2021kgd,Leljak:2021vte}) weaken this power
significantly. As recently reemphasized by us \cite{Buras:2021nns} this is in particular the case   of the branching ratios for K-meson decays 
  and the parameter $\varepsilon_K$ that exhibit stronger $|V_{cb}|$ dependences than rare $B$ decay branching ratios and the $\Delta M_{s,d}$ mass differences.
 
To cope with this difficulty one can consider
  within the SM suitable ratios of  two properly chosen observables
  so that the dependences on $\vcb$ and $\vub$ are eliminated
  \cite{Buras:2003td,Bobeth:2021cxm,Buras:2021nns}. While in \cite{Buras:2003td,Bobeth:2021cxm} $B$ physics observables were considered, our analysis in
  \cite{Buras:2021nns} was dominated by the $K$ system and its
  correlation with rare $B$ decays and $B_{s,d}^0-\bar B_{s,d}^0$ mixing.
  In this manner we could construct 16 $\vcb$ independent ratios that
  were either independent of the CKM parameters or only dependent
  on the angles $\beta$ and $\gamma$, that can be determined in tree-level
  processes. Having one day precise experimental values for the ratios
  in question and also precise values on $\beta$ and $\gamma$ will hopefully allow one to identify particular pattern of deviations from SM expectations
  independently of $\vcb$ and $\vub$ pointing towards a particular extension of the
  SM.

  This note summarizes the main results of \cite{Buras:2021nns} and 
is arranged as follows. In Section~\ref{sec:2} we will simply list
  the 16 ratios in question and briefly discuss their first implications. In Section~\ref{sec:2a} we report on inconsistencies between the determinations of $\vcb$ from $\varepsilon_K$, $\Delta M_d$ and $\Delta M_s$.
  In Section~\ref{sec:3}, combining these ratios with the assumption
  of no NP contributions to $|\varepsilon_K|$, $\Delta M_s$, $\Delta M_d$ and $S_{\psi K_S}$, we will present SM predictions for all branching ratios considered
  by us that are most accurate to date. Our note closes with a short outlook
  in Section~\ref{sec:4}.

  \boldmath
  \section{$\vcb$ and $\vub$ Independent Ratios}\label{sec:2}
  \unboldmath
As four basic CKM parameters we will use
\be\label{4CKM}
\boxed{\lambda=\vus,\qquad \vcb, \qquad \beta, \qquad \gamma}
\ee
with $\beta$ and $\gamma$ being two angles in the UT.
Their determination from mixing induced CP-asymmetries in tree-level $B$ decays {and using other tree-level strategies} is presently theoretically
cleaner than the determination of $\vub$.  A recent review of such determinations of $\beta$ and $\gamma$
can be found in \cite{Buras:2020xsm,Descotes-Genon:2017thz,Cerri:2018ypt}.

The $\vcb$ independent ratios constructed by us have
the general power-like structure
\be\label{criticalform}
  R_i=C_i[\sin\gamma]^{p_1}[\sin\beta]^{p_2},
  \ee
with the coefficients $C_i$ either being constants or being very weakly dependent on $\beta$ and $\gamma$. {Whenever the ratios are not already in the form of (\ref{criticalform}), we derive an approximate power-law expression, with non-integer exponents. The latter are indeed fitted to describe as power-law functions of parameters some more complicated exact expressions, with the best possible accuracy, which is $\lesssim 2\%$.} The dependence on $\vus$ is negligible, the angle $\beta$ is already known from $S_{\psi K_S}$ asymmetry with respectable precision and
  there is a significant progress by the LHCb collaboration on the determination of $\gamma$ from tree-level strategies \cite{LHCb:2021dcr}:
  \be\label{betagamma}
  \boxed{\beta=(22.2\pm 0.7)^\circ, \qquad  \gamma = (65.4^{+3.8}_{-4.2})^\circ \,.}
  \ee
  Moreover, in the coming years the determination of $\gamma$ by the LHCb and Belle II collaborations
  should be significantly improved so that precision tests of the SM using our
  strategy will be possible.

  The 16 $\vcb$ independent ratios in question are given as follows
\be
\boxed{R_0(\beta)=\frac{\mathcal{B}(\kpn)}{\mathcal{B}(\klpn)^{0.7}}=(2.03\pm 0.08)\times 10^{-3}\left[\frac{\sin 22.2^\circ}{\sin \beta}\right]^{1.4} 
={(2.03\pm 0.11)}\times 10^{-3}\,.
}
\label{eq:R0}
\ee
    \be\label{SR1}
\boxed{R_{\rm SL}=\frac{\BR(\ksm)_{\rm SD}}{\BR(\klpn)}=1.55\times 10^{-2}\,\left[\frac{\lambda}{0.225}\right]^2
\left[\frac{Y(x_t)}{X(x_t)}\right]^2\,,}
\ee
where $X(x_t)$ and $Y(x_t)$ are well known one-loop functions
with $x_t=m_t^2/M_W^2$ \cite{Buras:2020xsm}. Then, we have
\be\label{R12}
 \boxed{R_1(\beta,\gamma)=\frac{\mathcal{B}(\kpn)}{\left[{\overline{\mathcal{B}}}(B_s\to\mu^+\mu^-)\right]^{1.4}},\qquad
   R_2(\beta,\gamma)=\frac{\mathcal{B}(\kpn)}{\left[{\mathcal{B}}(B_d\to\mu^+\mu^-)\right]^{1.4}}.}
   \ee
 In particular the ratio $R_1$, implying the correlation in Fig.~\ref{fig:3abis}  should be of interest in 
the coming years due to  
the improved  measurement of $\kpn$ by NA62, of $B_s\to\mu^+\mu^-$ by LHCb, CMS and ATLAS and of $\gamma$ by LHCb {and Belle II}. Explicity we
have
\begin{align}
R_1(\beta,\gamma) &= {(55.24\pm 2.48)} \left[\frac{\sin\gamma}{\sin 67^\circ}\right]^{1.39} \left[\frac{G(22.2^\circ,67^\circ)}{G(\beta,\gamma)}\right]^{2.8}
\left[\frac{230.3\mev}{F_{B_s}}\right]^{2.8},\label{master1}
  \end{align}
\begin{align}
R_2(\beta,\gamma) &= {(8.29\pm 0.40)}\times 10^3 \left[\frac{\sin (67^\circ)}{\sin\gamma}\right]^{1.41} \left[\frac{190.0\mev}{F_{B_d}}\right]^{2.8}\label{master2},
\end{align}
where $F_{B_d}$ and $F_{B_s}$ are weak decay constants and
\be\label{vts}
G(\beta,\gamma)=
1 +\frac{\lambda^2}{2}(1-2 \sin\gamma\cos\beta)\,.
\ee
Furthermore, there are also the following $\vcb$ independent ratios
\be\label{R34}
  \boxed{R_3(\beta,\gamma)=\frac{\mathcal{B}(\klpn)}{\left[{\overline{\mathcal{B}}}(B_s\to\mu^+\mu^-)\right]^{2}},\qquad
   R_4(\beta,\gamma)=\frac{\mathcal{B}(\klpn)}{\left[{\mathcal{B}}(B_d\to\mu^+\mu^-)\right]^{2}}\,,}
   \ee
   which are given by
{\be\label{R3}
R_3(\beta,\gamma)=(2.17\pm 0.09)\times 10^6\left[{\frac{\sin\gamma\sin\beta}{\sin (67^\circ)\sin (22.2^\circ)}}\right]^2 \left[\frac{G(22.2^\circ,67^\circ)}{G(\beta,\gamma)}\right]^4\left[\frac{230.3\mev}{F_{B_s}}\right]^4\,,
\ee
\be\label{R4}
R_4(\beta,\gamma)=(2.79\pm 0.13)\times 10^{9}
  \left[{\frac{\sin (67^\circ)}{\sin\gamma}\frac{\sin\beta}{\sin (22.2^\circ)}}\right]^2\left(\frac{190.0\mev}{F_{B_d}}\right)^4.
  \ee}
In view of future improved results from NA62 and Belle II, of particular interest 
are the ratios
\be\label{R56}
\boxed{R_5(\beta,\gamma)=\frac{\mathcal{B}(\kpn)}{\left[\mathcal{B}(B^+\to K^+\nu\bar\nu)\right]^{1.4}},\qquad
R_6(\beta,\gamma)=\frac{\mathcal{B}(\kpn)}{\left[\mathcal{B}(B^0\to K^{0*}\nu\bar\nu)\right]^{1.4}}.}
\ee
Explicitly 
{\be\label{R5}
R_5(\beta,\gamma)={(2.69\pm0.51)}\times 10^{-3}
\left[\frac{\sin\gamma}{\sin 67^\circ}\right]^{1.39}\left[\frac{G(22.2^\circ,67^\circ)}{G(\beta,\gamma)}\right]^{2.8},
\ee
\be\label{R6}
R_6(\beta,\gamma)={(9.07\pm1.23)}\times 10^{-4}
\left[\frac{\sin\gamma}{\sin 67^\circ}\right]^{1.39}\left[\frac{G(22.2^\circ,67^\circ)}{G(\beta,\gamma)}\right]^{2.8}\,,
\ee}
with $G(\beta,\gamma)$ defined in (\ref{vts}). Next, in view of future LHCb and Belle II measurements, of particular interest are the ratos
\be\label{R78}
\boxed{R_7=\frac{\mathcal{B}(B^+\to K^+\nu\bar\nu)}{{\overline{\mathcal{B}}}(B_s\to\mu^+\mu^-)},\qquad
R_8=\frac{\mathcal{B}(B^0\to K^{*0}\nu\bar\nu)}{{\overline{\mathcal{B}}}(B_s\to\mu^+\mu^-)}}
    \ee
    with 
    \be\label{R7SM}
    (R_7)_\text{SM}={(1.20\pm 0.17)}\times 10^{3}\left(\frac{230.3\mev}{F_{B_s}}\right)^2\,,
    \ee
     \be\label{R8SM}
    (R_8)_\text{SM}={(2.62\pm 0.25)}\times 10^{3}\left(\frac{230.3\mev}{F_{B_s}}\right)^2\,.
     \ee

Turning our attention to $\varepsilon_K$ and $\Delta M_{s.d}$ we  define two $\vcb$ independent ratios
  \be
  \boxed{R_9(\beta,\gamma)=\frac{ |\varepsilon_K|}{(\Delta M_d)^{{1.7}}},\qquad
  R_{10}(\beta,\gamma)=\frac{ |\varepsilon_K|}{(\Delta M_s)^{{1.7}}}\, .}
  \ee
  The explicit expressions for them read
  \be\label{R9}
  {R_9(\beta,\gamma)=6.405\times 10^{-3}\, {\rm ps^{1.7}}\left(\frac{\sin 67^\circ}{\sin \gamma}\right)^{1.73}\left(\frac{\sin \beta}{\sin 22.2^\circ}\right)^{0.87}
\bar{R}^\epsilon_d },
    \ee
\be\label{R10}
  {R_{10}(\beta,\gamma)=1.516\times 10^{-5}\, {\rm ps^{1.7}}\left(\frac{\sin \gamma}{\sin 67^\circ}\right)^{1.67}\left(\frac{\sin \beta}{\sin 22.2^\circ}\right)^{0.87}\left(\frac{G(22.2^\circ,67^\circ)}{G(\beta,\gamma)}\right)^{3.4}\,
\bar{R}^\epsilon_s }.
  \ee
  Here
  {\be
  \bar{R}^\epsilon_d=\left( 
\frac{214.0\mev}{\sqrt{\hat B_{B_d}}F_{B_d}}\right)^{3.4}
\left(\frac{2.307}{S_0(x_t)}\right)^{1.7} 
\left(\frac{0.5521}{\eta_B}\right)^{1.7}\, ,
  \ee
  \be
   \bar{R}^\epsilon_s=\left( 
\frac{261.7\mev}{\sqrt{\hat B_{B_s}}F_{B_s}}\right)^{3.4}
\left(\frac{2.307}{S_0(x_t)}\right)^{1.7} 
\left(\frac{0.5521}{\eta_B}\right)^{1.7} \, .
\ee

 Finally, there are also the following four ratios
\be\label{R11}
  \boxed{R_{11}(\beta,\gamma)=\frac{\mathcal{B}(\kpn)}{|\varepsilon_K|^{0.82}}=(1.31\pm0.05)\times 10^{-8}{\left(\frac{\sin\gamma}{\sin 67^\circ}\right)^{0.015}\left(\frac{\sin 22.2^\circ}{\sin \beta}\right)^{0.71},  }            }
  \ee
  \be\label{R12a}
\boxed{R_{12}(\beta,\gamma)=\frac{\mathcal{B}(\klpn)}{|\varepsilon_K|^{1.18}}=(3.87\pm0.06)\times 10^{-8}
    {\left(\frac{\sin\gamma}{\sin 67^\circ}\right)^{0.03}\left(\frac{\sin\beta}{\sin 22.2^\circ}\right)^{0.9{8}},}}
  \ee
  \be\label{CMFV6}
\boxed{R_q=\frac{\mathcal{B}(B_q\to\mu^+\mu^-)}{\Delta M_q}= 4.291\times 10^{-10}\ \frac{\tau_{B_q}}{\hat B_q}\frac{(Y_0(x_t))^2}{S_0(x_t)},\qquad q=d,s\,.}
\ee
The first two of these formulae express explicitly the fact that combining
  on the one hand $\kpn$ and $\varepsilon_K$ and on the other hand
  $\klpn$ and  $\varepsilon_K$ allows within the SM to determine
  to a very good approximation the angle $\beta$ independently of the value
  of $\vcb$ and $\gamma$. This is one of the most interesting new results of
  our paper. The CKM independence of $R_q$ has been pointed out already
  in \cite{Buras:2003td}.

There are two first interesting consequences of these $\vcb$ independent relations.
  First, combining the LHCb and Belle II results we find  \cite{Buras:2021nns}
    \be
    (R_7)_{\rm EXP}=(3.86\pm 1.48)\times 10^{3}.
       \ee
       The central value is by a factor of 3.2 larger than the SM prediction (\ref{R7SM}) but due to large error in the experimental $B^+\to K^+\nu\bar\nu$ branching ratio the tension  is only at $1.8\sigma$.

       The second result is for $R_s$ in (\ref{CMFV6}) for which one
       finds \cite{Bobeth:2021cxm}
       \be
       (R_s)_{\rm SM}=(2.042^{+0.083}_{-0.058})\times 10^{-10}{\rm ps},\qquad
       (R_s)_{\rm EXP}=(1.61^{+0.19}_{-0.17})\times 10^{-10}{\rm ps},
       \ee
       that is a $2.1~\sigma$ tension.

       {While the ratios presented here are independent of $\vcb$ and
         $\vub$ they depend on the hadronic parameters which enter in particular
         the ratios involving  $|\varepsilon_K|$, $\Delta M_d$ and $\Delta M_s$.
       This brings us to the next important topic.}
       \boldmath
\section{$\vcb(\beta,\gamma)$ from $|\varepsilon_K|$, $\Delta M_d$ and $\Delta M_s$}\label{sec:2a}
\unboldmath
When one day the angle $\gamma$ will be measured precisely we will be
able to answer the question whether all these SM predictions for the ratios in question can be satisfied
by the data simultaneously. We have just seen two possible tensions.
But in fact, we can do more already today. Namely, we can ask the question,
whether the SM can describe the very precise data for 
$\varepsilon_K$, $\Delta M_s$ and $\Delta M_d$ with the same values of $\vcb$,
$\beta$ and $\gamma$. Also, theory in this three cases is in a good shape due
to progress from LQCD groups over many years and in the case of $\varepsilon_K$
due to the progress in \cite{Brod:2019rzc}.

{To answer this question  we considered in our paper the following hadronic parameters
  \be\label{CBAJB}
  F_{B_d}\sqrt{\hat{B}_{B_d}}= 214.0(39)\,{\rm MeV},\quad F_{B_s}\sqrt{\hat{B}_{B_s}}= 261.7(38)\,{\rm MeV}\,
  \ee
  in the case of $\Delta M_{d,s}$ and $\hat B_K=0.7625(97)$ in the case of
  $\varepsilon_K$.} {The latter one is the PDG average, while the values in   (\ref{CBAJB}) are the averages of $2+1$ and $2+1+1$ LQCD results quoted by  PDG obtained
  in  \cite{Bobeth:2021cxm}. The result of this exercise is shown 
  in Fig.~\ref{fig:5}.}

We find  that {with the chosen hadronic parameters} it is not possible to obtain simultaneous
    agreement with 
  the data on $\varepsilon_K$, $\Delta M_d$, $\Delta M_s$ and
  $S_{\psi K_S}$, used to determine $\beta$, within the SM independently of the value of $\vcb$ and $\gamma$.
  While these tensions are  still moderate they could hint for some NP at work. In this   context a precise measurement of $\gamma$ and the improvements on hadronic parameters will be important. The consequences of these findings are discussed in detail in \cite{Buras:2021nns}. But the
  message is clear. The ratios involving $\varepsilon_K$, $\Delta M_d$, $\Delta M_s$, that is
  \be
  R_9,\qquad R_{10},\qquad  R_{11}, \qquad R_{12}, \qquad R_d, \qquad  R_s,
  \ee
  are not expected to agree simultaneously with the future data {for the
    set of hadronic parameters in (\ref{CBAJB})}. This
  lead us to propose a different strategy for SM predictions for rare $K$ and $B$ decays summarized   in the rest of this writing.
   \begin{figure}[t!]
\centering%
\includegraphics[width=0.70\textwidth]{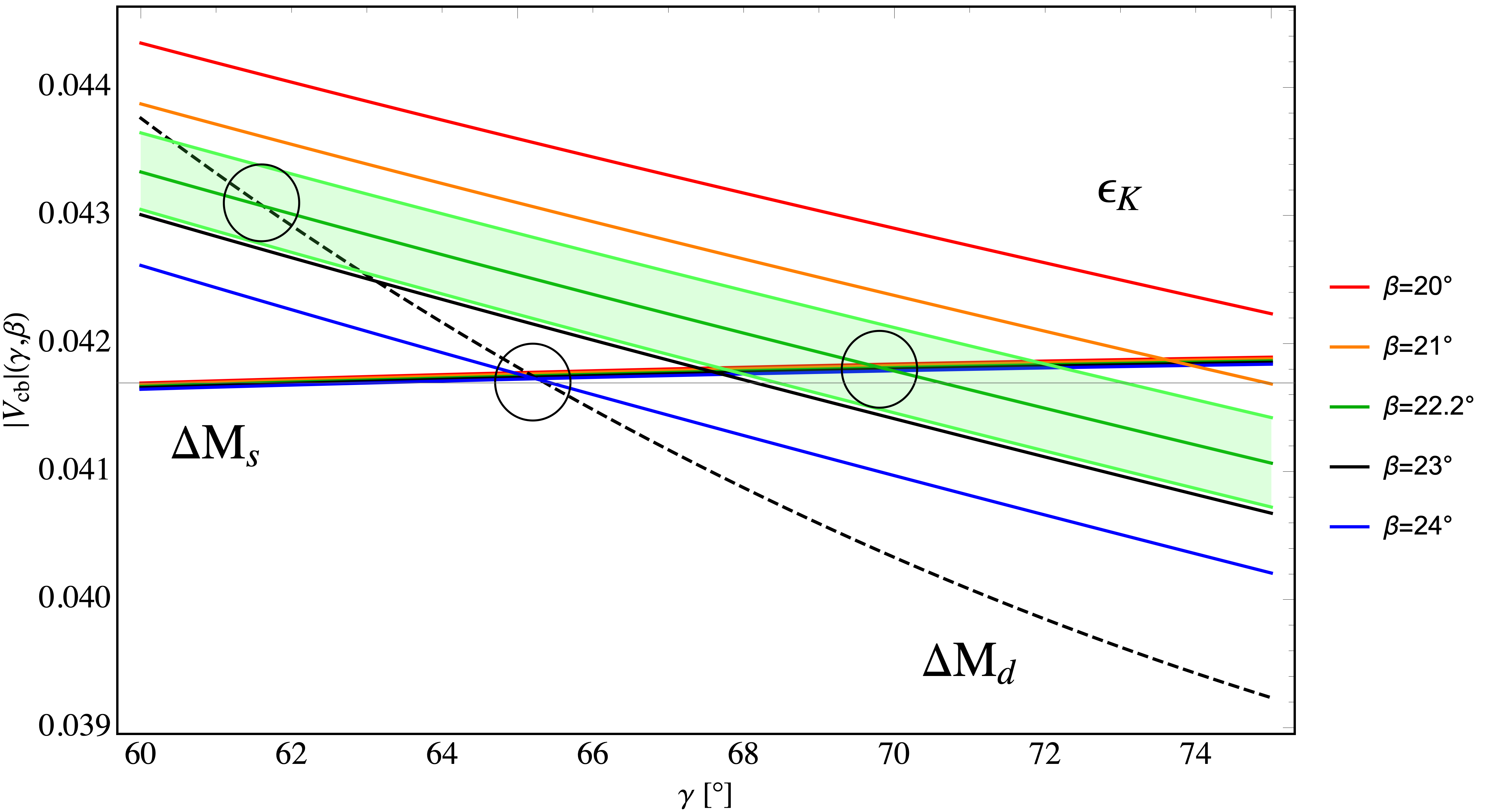}%
\caption{\it {The values of $\vcb$ extracted from $\varepsilon_K$, $\Delta M_d$ and  $\Delta M_s$ as functions of $\gamma$ for different values of $\beta$. {$\vcb$ extracted from $\Delta M_d$ is independent of  $\beta$.}} 
\label{fig:5}}
\end{figure}

\boldmath
\section{{Improved SM Predictions for rare $K$ and $B$ Decays}}\label{sec:3}
\unboldmath
While the ratios listed above are 
useful in the context of the testing of the SM and in the search for NP, they are
 not as interesting as
  the observables themselves. Therefore, assuming in addition no
  NP in $\varepsilon_K$, $\Delta M_d$ and $\Delta M_s$  and in the mixing induced   CP-asymmetry $S_{\psi K_S}$, these ratios allowed to obtain 
  $\vcb$ independent SM predictions for a number
  of branching ratios \cite{Buras:2021nns}. As these four quark mixing observables are very   precisely measured and theoretically rather clean, the resulting predictions obtained in this manner turned out to be the most precise to date.
  We report on these predictions in Table~\ref{tab:SMBRs}.
 \begin{table}
\centering
\renewcommand{\arraystretch}{1.4}
\resizebox{\columnwidth}{!}{
\begin{tabular}{|ll||ll|}
\hline
Decay 
& Branching Ratio
& Decay
&  Branching Ratio
\\
\hline \hline
 $\kpn$ & $(8.60\pm 0.42)\times 10^{-11}$ & $B_s\to\mu^+\mu^-$ & $(3.62^{+ 0.15}_{-0.10})\times 10^{-9}$
\\
 $\klpn$ & $(2.94\pm 0.15)\times 10^{-11}$ & $B_d\to\mu^+\mu^-$ & $(0.99^{+ 0.05}_{-0.03})\ \times 10^{-10}$
\\
$\ksm$ & {$(1.85\pm 0.10)\times 10^{-13}$} &  $B^+\to K^+\nu\bar\nu$ & $(4.45\pm 0.62)\times 10^{-6}$
\\
& &  $B^0\to K^{0*}\nu\bar\nu$ &$(9.70\pm 0.92)\times 10^{-6}$
\\
\hline
\end{tabular}
}
\renewcommand{\arraystretch}{1.0}
\caption{\label{tab:SMBRs}
  \small
  \small
  Present most accurate $\vcb$ independent SM estimates  of the branching ratios considered in the paper. The $\gamma$ dependence is either very small or absent. See   \cite{Buras:2021nns} for details.
}
\end{table}
  
 {However, in view of the inconsistencies in different determinations
   of $\vcb$ seen in Fig.~\ref{fig:5}
to obtain these results we did not perform on purpose the usual global fit of observables which would require in addition the input
 on $\vcb$ from tree-level decays in contradiction with the main strategy
 of our paper. 
 We concluded therefore that it would be a bad idea to assume, as done in global fits,} that
 NP is absent simultaneously in $\varepsilon_K$, $\Delta M_d$, $\Delta M_s$ and   $S_{\psi K_S}$.
Therefore, to obtain SM predictions for  rare Kaon decays we only assumed
the absence of NP in $\varepsilon_K$ and $S_{\psi K_S}$. To obtain predictions
for $B_s\to \mu^+\mu^-$ and  $B_d\to \mu^+\mu^-$ we assumed, following \cite{Buras:2003td}, the absence of NP in $\Delta M_s$ and in ($\Delta M_d$,$S_{\psi K_S}$), respectively but
not simultaneously. In our view this strategy for finding SM predictions is presently more powerful than any global fit which would include decays like $B\to K\mu^+\mu^-$, $B\to K^*\mu^+\mu^-$ and $B_s\to\phi\mu^+\mu^-$ that seem to exhibit significant contributions from NP. Moreover, these {\em local}  assumptions of the absence of NP, with the goal to
obtain SM predictions, are weaker than made in {\em global} fits. {But one should
be aware of the fact that the SM predictions in Table~\ref{tab:SMBRs} {that are based on hadronic parameters in (\ref{CBAJB})} are likely
not be true simultaneously in view of the results in Fig.~\ref{fig:5}. But this is precisely
what we are after!}

\begin{figure}[t]
\centering%
\includegraphics[width=0.455\textwidth]{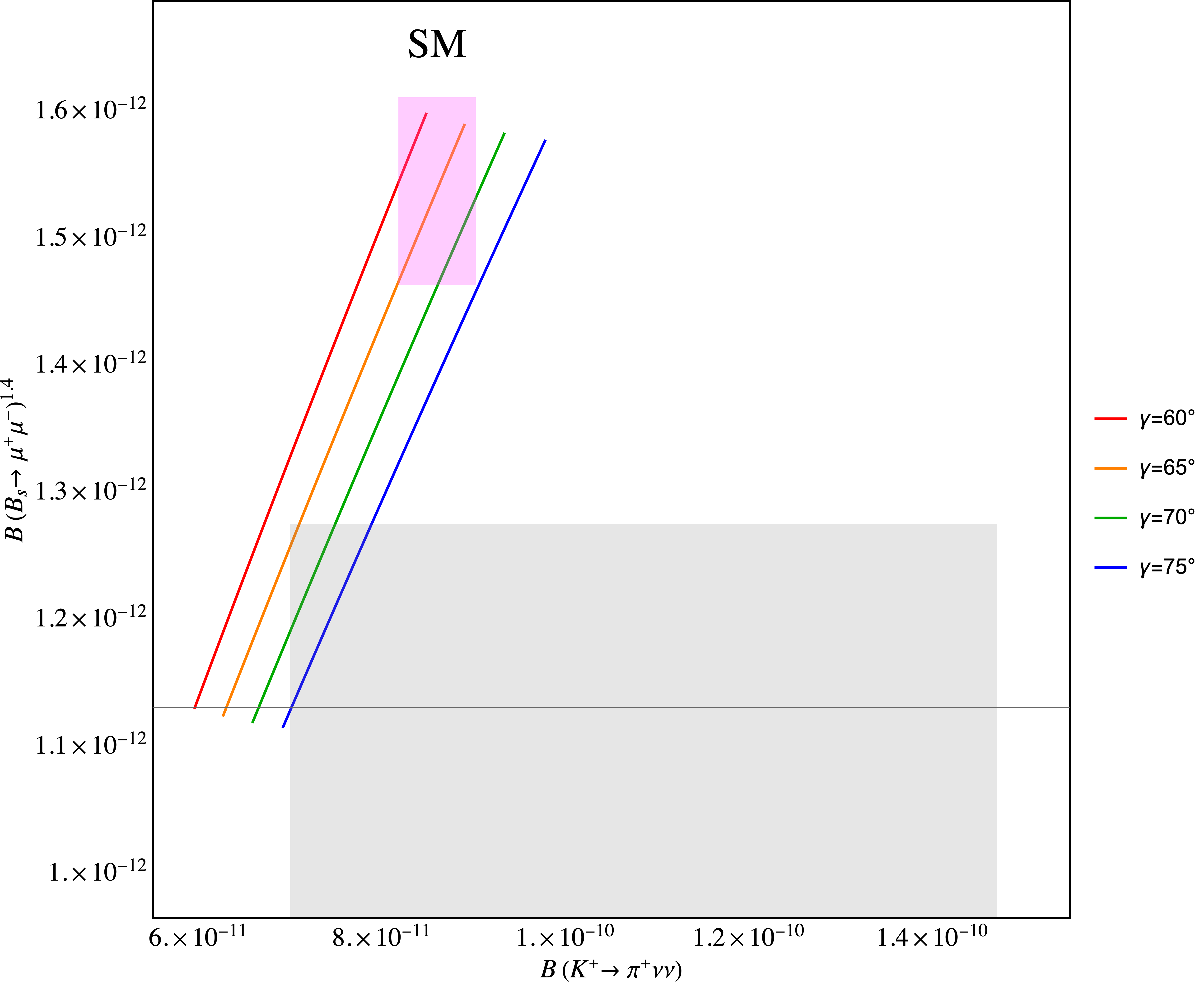}
\includegraphics[width=0.45\textwidth]{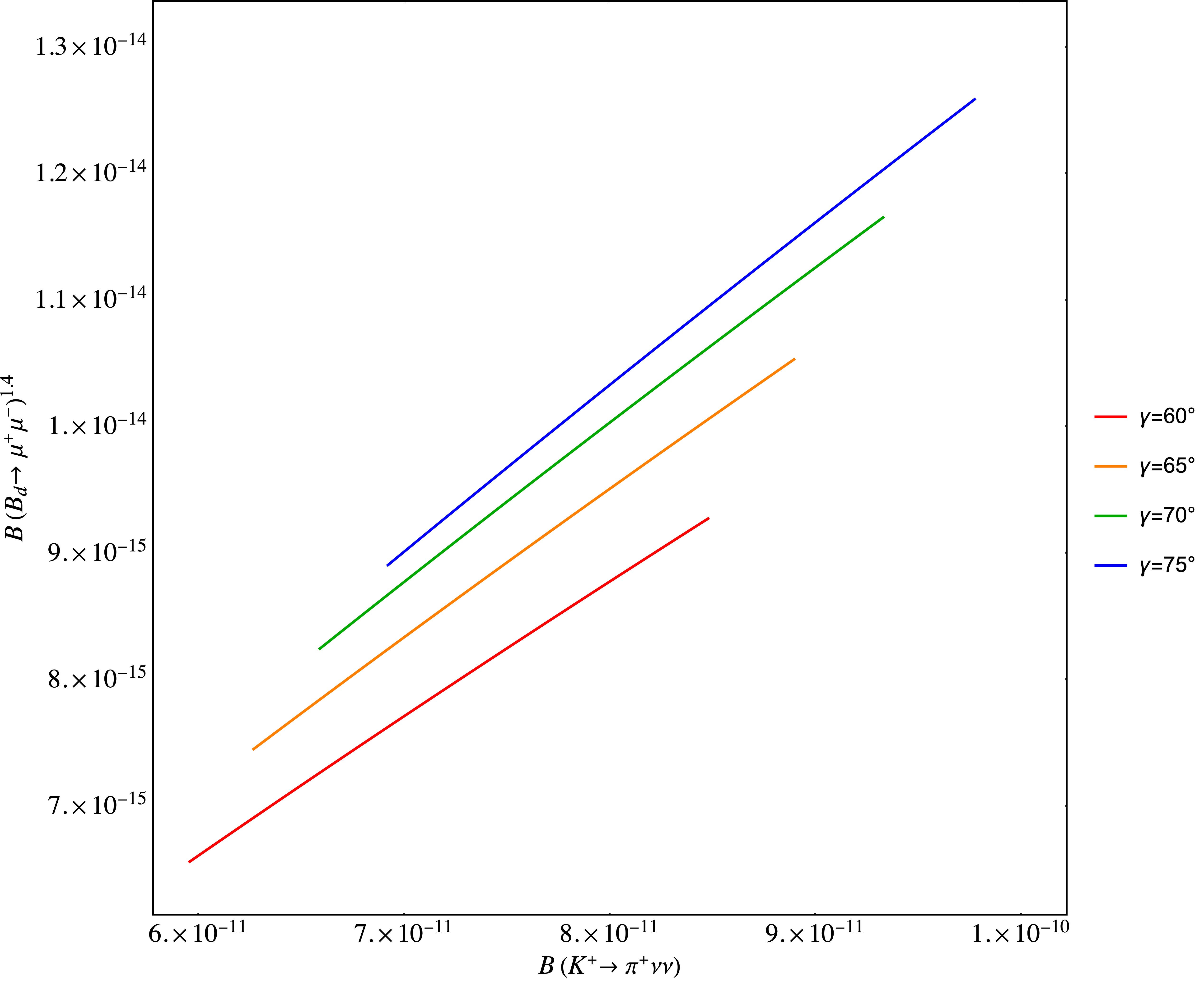}%
\caption{\it {The correlations of $\mathcal{B}(\kpn)$ with $\overline{\mathcal{B}}(B_s\to\mu^+\mu^-)^{1.4}$ (left panel) and with ${\mathcal{B}}(B_d\to\mu^+\mu^-)^{1.4}$ (right panel) as given in (\ref{master1}) and (\ref{master2}), for different  values of $\gamma$ within the SM.
    The ranges of branching ratios correspond to $38 \leq |V_{cb}|\times 10^{3} \leq 43$ {and $20^\circ\leq\beta \leq 24^\circ$.}} The SM area corresponds to the one in Table~\ref{tab:SMBRs}.
    The gray area represents the
present experimental situation.} \label{fig:3abis}
\end{figure}

\section{Summary and Outlook}\label{sec:4}
We have summarized the main results of our recent paper \cite{Buras:2021nns},
where following and extending significantly the strategies of
\cite{Buchalla:1994tr,Buras:1994rj,Buchalla:1996fp,Buras:2002yj,Buras:2015qea,Buras:2003td,Bobeth:2021cxm,Blanke:2018cya}, we have proposed to search for NP in rare Kaon and $B$-meson decays without the necessity
  to choose the values of the CKM elements $\vcb$ and $\vub$, that introduce
  presently large parametric uncertainties in the otherwise theoretically
  clean decays $\kpn$, $\klpn$, $\ksm$, $B_{s,d}\to\mu^+\mu^-$, $B\to K(K^*)\nu\bar\nu$, in the parameter $\varepsilon_K$ and $\Delta M_{s,d}$.
  The 16 $\vcb$ independent ratios of branching ratios and mixing observables,
  depending only 
  on the angle $\beta$ and $\gamma$ in the UT, will certainly play an important role
  in the search for NP as soon as the branching ratios in question and the angle
  $\gamma$ will be   precisely measured. Several plots of these ratios
  and also of correlations between various branching ratios can be found in our paper. We present here only
  two correlations in Fig.~\ref{fig:3abis} which could turn out in the coming
  years to be the smoking guns of NP. {{For the set of hadronic parameters in (\ref{CBAJB}) we} have also identified inconsistencies between the determinations of $\vcb$ from $\varepsilon_K$, $\Delta M_d$ and $\Delta M_s$. This led us to propose a novel strategy to obtain SM predictions for
    rare $K$ and $B$ decay branching ratios which is based on {\em locality}
    rather than {\em globality} as explained in Section~\ref{sec:2a}.}

\section*{ Acknowledgements}

\noindent
{We would like to thank Pietro Baratella, Jean-Marc G{\'e}rard and Peter Stangl for discussions.}
A.J.B acknowledges financial support from the Excellence Cluster ORIGINS,
funded by the Deutsche Forschungsgemeinschaft (DFG, German Research Foundation), 
Excellence Strategy, EXC-2094, 390783311. E.V. has been partially funded by the Deutsche Forschungs-gemeinschaft (DFG, German Research Foundation) under Germany's Excellence Strategy- EXC-2094 - 390783311, by the Collaborative Research Center SFB1258 and the BMBFgrant  05H18WOCA1  and  thanks  the  Munich  Institute  for  Astro-  and  Particle  Physics(MIAPP) for hospitality.

\bibliographystyle{JHEP}
\bibliography{Bookallrefs}
\end{document}